\documentclass[pre,twocolumn,preprintnumbers,amsmath,amssymb,tightenlines,epsfig]{revtex4}

\usepackage{graphicx}
\usepackage{dcolumn}
\usepackage{bm}
\usepackage{epsfig}
\usepackage{stmaryrd}
\usepackage[usenames,dvipsnames]{color}

\usepackage[utf8]{inputenc}

\def\beq{\begin{align}}
\def\eeq{\end{align}}

\newcommand{\rref}[1]{Ref.~\cite{#1}}
\newcommand{\fref}[1]{FIG.~\ref{#1}}

\newcommand{\frefp}[2]{FIG.~\ref{#1}#2}

\newcommand{\eref}[1]{Eq.~(\ref{#1})}

\newcommand{\sref}[1]{section~\ref{#1}}

\newcommand{\cref}[1]{chapter~\ref{#1}}

\newcommand{\Cref}[1]{Chapter~\ref{#1}}
\newcommand{\tref}[1]{table~\ref{#1}}

\newcommand{\bref}[1]{(\ref{#1})}

\newcommand{\bra}[1]{\langle\,{#1}\, |}
\newcommand{\ket}[1]{|\,{#1}\,\rangle}

\usepackage[normalem]{ulem}

\newcommand{\siteenergy}{\varepsilon}
\newcommand{\Eoffset}{\epsilon}

\graphicspath{../}

\begin{document}

\title{Open quantum system parameters from molecular dynamics}
\author{Xiaoqing Wang}
\affiliation{Max-Planck-Institut f\"ur Physik komplexer Systeme, N\"othnitzer Str.~38, D-01187 Dresden, Germany}
\author{Gerhard Ritschel}
\affiliation{Max-Planck-Institut f\"ur Physik komplexer Systeme, N\"othnitzer Str.~38, D-01187 Dresden, Germany}
\author{Sebastian W{\"u}ster}
\affiliation{Max-Planck-Institut f\"ur Physik komplexer Systeme, N\"othnitzer Str.~38, D-01187 Dresden, Germany}
\author{Alexander Eisfeld}
\email{eisfeld@mpipks-dresden.mpg.de}
\affiliation{Max-Planck-Institut f\"ur Physik komplexer Systeme, N\"othnitzer Str.~38, D-01187 Dresden, Germany}

\begin{abstract}
We extract the site energies and spectral densities of the Fenna-Matthews-Olson (FMO) pigment protein complex of green sulphur bacteria from 
simulations of molecular dynamics combined with energy gap calculations. Comparing four different combinations of methods, we investigate the origin of  
quantitative differences regarding site energies and spectral densities obtained previously in the literature. We find that different forcefields for molecular dynamics  and 
varying local energy minima found by the structure relaxation yield significantly different results. Nevertheless, a picture averaged over these variations is in good agreement with experiments and some other theory results. Throughout, we discuss how vibrations external- or internal to the pigment molecules enter the extracted quantities differently and can be distinguished. Our results offer some guidance to set up more computationally intensive  calculations for a precise determination of spectral densities in the future. These are 
 required to determine absorption spectra as well as transport properties of light-harvesting complexes. 
\end{abstract}

\maketitle

\section{Introduction}
\label{introduction}

Understanding the influence of the protein environment and exciton phonon  coupling on the electronic excitation transfer in natural light harvesting systems is one of the great challenges of photosynthesis research. 
Unfortunately, a full quantum mechanical treatment of even small photosynthetic complexes seems out of reach in the near future. 
Instead, one seeks approaches that keep as much microscopic detail as possible and are still tractable with reasonable computational effort.
One viable approach is to treat nuclear motion classically using molecular dynamics (MD) and excitation transfer quantum mechanically, using methods for open quantum systems. The latter require input parameters that are  extracted from the MD simulation.

\begin{figure}[htb]
\includegraphics[width=0.99\columnwidth]{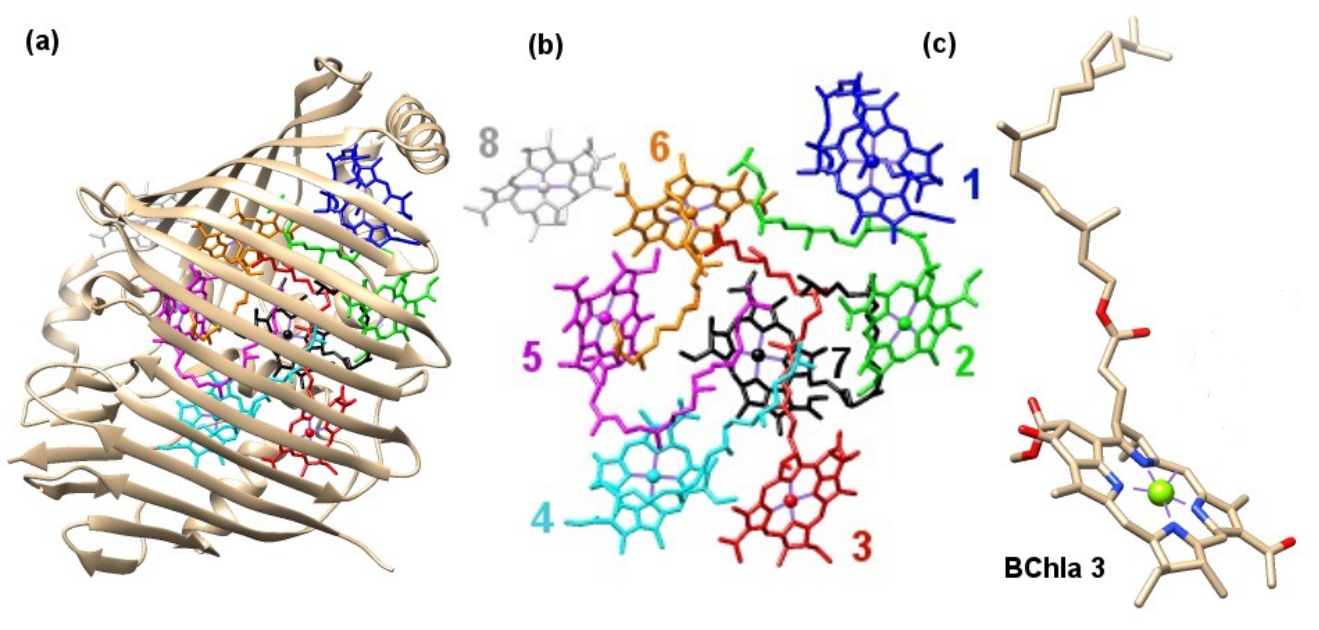}
\caption{\label{fig:FMO_sketch}(a) Sketch of the FMO monomer including the protein (brown stripes). (b)  FMO monomer without protein, showing the numbering of the BChl molecules. (c) Structure of an individual BChl molecule.}
\end{figure}

In this article we compare different implementations of MD simulations in combination with different ways of extracting open quantum system parameters from these, to understand large deviations among parameters found in this manner in the literature \cite{OlJaLi11_8609_,OlStSc11_1771_,RiMoMa13_5510_,ShReVa12_649_,VaEiAs12_224103_}.
 As model system we consider one subunit of the trimeric Fenna-Matthews-Olson (FMO) complex of {\it C. tepidum}, shown in \fref{fig:FMO_sketch}, since it is most widely studied and with only eight Bacteriochlorophyll (BChl) pigment molecules per subunit quite small and tractable.

To investigate excitation energy transfer or optical spectra, knowledge of excited state dynamics is required.
One has to consider the presence of at least one electronic excitation in the photosynthetic complex, which can migrate among its pigment sites and even be shared coherently between several of them.
These processes in turn also react back on the internal degrees of freedom of the molecules  and the protein, since a change in electronic state induces a redistribution of charge-density and leads to vibrations. For these reasons, excited states pose serious problems for MD simulations.
In contrast, ground state MD is capable of simulating the thermal motion of systems with tens of thousands of atoms with reasonable resources, and can therefore be used for quite large light harvesting complexes.  
Fortunately, one can obtain some of the essential parameters determining the electronic excited state dynamics also from MD simulations in the ground state.

 Typically, electronic excited state dynamics is tackled using open quantum system methods, where the relevant electronic levels of the BChls span the system of interest and the remaining degrees of freedom form the quantum environment. These methods require two essential inputs: 
(i) A system Hamiltonian, which contains the so-called site energies of the BChls on the diagonal and the off-diagonal elements of which describe exciton transfer couplings between the BChls.
For simplicity one neglects higher- or multiply excited electronic states of the BChls, thus reducing the problem to the so-called one-exciton manifold.
 (ii)  Bath spectral densities, which describe the couplings of the BChl electronic excitations to  baths of harmonic oscillators. 
The question is then how to obtain these quantities, which require knowledge of the electronically excited state, from the molecular dynamics simulations.

The method discussed here, is to calculate the (time-dependent) electronic transition energy of the individual BChls, 
which are often referred to as 'energy gap functions', along a ground state MD trajectory.
The transition energies of the BChls depend both on the geometry of the BChl and also on the conformation of the surrounding protein, which creates an electric field at the BChl locations.
The average of the transition energy of each BChl along the trajectory is then taken as the site energy of the respective BChl.
Similarly, the (time dependent) transition dipole-dipole couplings between pairs of BChls are calculated along the MD trajectory. 
The main difficulty is to obtain the spectral densities describing the influence of the quantum environment on the system Hamiltonian. 
Here, as in most previous works, we focus on the spectral densities of the transition energies.
To this end, a {\it classical} 'gap correlation' function is calculated from the transition energies of each BChl. Then, using semi-classical {\it a posteriori} corrections, the quantum correlation function of the environment is constructed, see Refs.~\cite{DaKoKl02_031919_,VaEiAs12_224103_}.

This method has been applied in recent years to various  light harvesting complexes \cite{DaKoKl02_031919_,OlKl10_12427_,OlJaLi11_8609_,OlStSc11_1771_,OlStSc11_758_,ShReVa12_649_,VaEiAs12_224103_,RiMoMa13_5510_,JuCuMe14_3138_,AgStOl14_3131_,janosi:LH2_QMMM,JiZhLi12_1164_}. The electronic system Hamiltonians extracted from these calculations typically agree reasonably well with that extracted from fitting simple exciton models to experimental spectra.
However, they show a large spread in parameters, even the energetic ordering of the site energies is not agreed upon by all calculations.
Furthermore, there are large deviations between the gap-energy correlation functions and resulting spectral densities obtained in different works and
their agreement with spectral densities extracted experimentally from fluorescence line narrowing techniques \cite{WePuPr00_5825_,RaeFr07_251_,AdRe06_2778_,raetsep:optspecclorpphyll} is at best qualitative and often very poor. Also fundamental questions about the applicability of the approach, such as  whether a harmonic model is appropriate, remain open and require more detailed investigations.

A crucial element of the  scheme is the calculation of transition energies along the trajectory. 
Typically, these are calculated with quantum chemical methods, such as time-dependent density functional theory (TDDFT) in Ref.~\cite{ShReVa12_649_} or semi-empirical methods in Refs.~\cite{DaKoKl02_031919_,OlStSc11_1771_,OlStSc11_758_,OlJaLi11_8609_,GaShYe13_3488_}. 
Mostly, these methods account for the influence of the protein on the pigments (here the BChls), by treating the protein environment as a collection of partial charges.

An alternative to quantum chemical calculations are classical electrostatic calculations. 
The latter gave promising results for site energies, calculated using the crystal structure, without including dynamics \cite{AdRe06_2778_}.
It has been suggested in Ref.~\cite{AdMueMa08_197_} to use this approach combined with MD simulations. 
Recently, this proposal has been implemented \cite{RiMoMa13_5510_}, calculating transition energies using the so-called {\it charge density coupling} (CDC) method~\cite{AdMueMa08_197_}. 
There, one calculates the transition energy from the interaction of the partial charges of the protein with 'transition charges' of the BChls.

As indicated above, the transition energy along a trajectory (i.e.~the energy gap function) plays a fundamental role for both, the determination of the site energies and spectral densities.
We will show in this article that the details of the molecular dynamics simulations (for example the force fields, or the initial condition) and the methods used to calculate the transition energies (classical versus quantum chemical, and choice of quantum chemical method) strongly influence the obtained energy gap functions and the derived quantities. 
 To make comprehensive checks on convergence and initial state preparation, we use computationally fast methods to calculate the electronic energies. 
Our results thus offer guidance in setting up the MD and energy calculation architecture for future applications of more advanced and computationally intensive methods.

The present work does not attempt a fully quantitative description of the FMO complex. Rather we wish to present an overview of the implications and limitations of the methods commonly employed for such a description. To facilitate this, we focus on the monomeric structure of the FMO complex. We include the 8'th BChl molecule despite it dissolving in MD simulations \cite{OlJaLi11_8609_}, since the underlying structure determination  \cite{BeFrNe04_274_} (PDB code 3BSD) is more recent than for the corresponding structures with 7 BChls.

The article is organized as follows: In section \ref{opensyst} we will define the Hamiltonian underlying the open system framework for which we seek to extract parameters from the MD simulations.
Then, in section \ref{method} details are given about the MD simulations, the calculations of the 'gap energies' and determination of open system model parameters from these. Section \ref{results} contains our results, grouped in basic properties of energy-gap trajectories (\ref{sec:trajects}), energy correlations (\ref{correlations}) and pigment spectral densities (\ref{specdens}). We summarize and interpret our results in the conclusion, \sref{conclusions}. A comprehensive collection of our data is provided in the supporting information \cite{supp:inf}.

\section{Open system approach, spectral densities}
\label{opensyst}
One often uses open system techniques to calculate optical- and transport properties of pigment protein complexes, as discussed in the introduction. To set the stage for these, consider a Hamiltonian divided into $H=H_{ex}+H_{vib}+H_{ex-vib}$. 
Here $H_{ex}$ is the system Hamiltonian containing the relevant electronic states for  the optical and transport properties. 
For linear spectra and excitation transfer it is sufficient to consider a restricted basis of states $\ket{n}=\ket{g}\cdots\ket{e}_n\cdots\ket{g}$, in which BChl $n$ is electronically excited to $\ket{e}$ while the other BChls are in their ground state $\ket{g}$. 
Then we have
\begin{equation}
\label{Hex}
H_{ex}=\sum_n \siteenergy_n \ket{n}\bra{n} + \sum_{nm} V_{nm} \ket{n}\bra{m}
\end{equation}
where $\siteenergy_n$ is the transition energy, i.e.~the energy difference between ground and excited state of BChl $n$, and $V_{nm}$ is the interaction associated with a transfer of excitation from BChl $m$ to $n$.  Throughout, indices $n,m = 1 \ldots 8$ label the BChl molecules as shown in Fig.~\ref{fig:FMO_sketch}(b).

The interaction has a strong dependence on the distance and the orientation of the BChl molecules.
In its simplest and most common form $V_{nm}$ is written in the point dipole-dipole interaction~\cite{AmVaGr00__} as 
\begin{equation}
\label{eq:point_dipol-dipol}
V_{nm}=f \frac{\mu_n\mu_m}{R_{nm}^3}[\vec{\mu}_n\cdot \vec{\mu}_m-3(\vec{\mu}_n\cdot \hat{R}_{nm})(\vec{\mu}_m\cdot \hat{R}_{nm})], 
\end{equation}
where $\vec{\mu}_n$ is the transition dipole vector of BChl $n$ and $\hat{R}_{nm}$ is the distance vector between the Mg atoms of BChls $n$ and $m$. 
The factor $f$ takes into account the dielectric environment, see \cite{AdRe06_2778_,Scholes_solventscreening} and references therein. 
We show Eq.~\bref{eq:point_dipol-dipol} for completeness, but do not consider interactions in this article. 

The environment is modelled as an (infinite) bath of harmonic oscillators $H_{vib}=\sum_{n=1}^8 H_{vib}^{(n)}=\sum_n\sum_j \hbar \omega_{nj} a^{\dagger}_{nj} a_{nj}$ and the system-environment interaction is taken as $H_{ex-vib}=\sum_{n=1}^8\sum_j \kappa_{nj} \ket{\pi_n}\bra{\pi_n} (a^{\dagger}_{nj} +a_{nj})$.
Note that here we have assumed that all BChls couple to independent baths.
This is probably a good approximation for internal vibrations of the BChls, while nuclear dynamics of the protein could simultaneously affect multiple BChls.
In \sref{correlations} we will discuss this point further.

 The relevant quantity to describe the effect of the bath on one BChl molecule is the  so-called spectral density,
\begin{equation}
J_n(\omega)=\sum_j |\kappa_{nj}|^2 \delta(\omega-\omega_{nj}),
\end{equation}
 which describes how strong the electronic transition of BChl $n$ couples to its oscillator $j$ with  frequency $\omega_{nj}$.
Typically, these spectral densities are considered to be  continuous functions of frequency. 

The SDs  enter  open quantum system approaches via the so-called bath-correlation functions (see e.g.\ Ref.~\cite{VaEiAs12_224103_,MaKue00__})
\begin{equation}
C_n(t)= \frac{1}{\pi}\int_0^{\infty} {\rm d}\omega \, J_n(\omega) \left[ \coth\left(\frac{\hbar \omega}{2 k_{\rm B}T}\right) \cos(\omega t) - i \sin(\omega t) \right]
\end{equation}
Here $n$ refers again to the respective BChl and $T$ is the temperature.

\section{Methods}
\label{method}
Since the system part and the environmental part of the model introduced in the previous section both arise from molecular quantum dynamics of the FMO protein, they could in principle be extracted from ab-initio modelling of this structure. 
However, as a full quantum description of the complex is out of the question, we resort to determine thermal classical (electronic ground state) motion of the complex using molecular dynamics simulations described in \sref{MD}. 
From the molecular coordinates thus obtained, we use semi-classical or quantum-chemical methods to obtain time-dependent energy differences between ground- and excited states of the BChls along the MD trajectory, see \sref{opensystparams}. From these, in turn, we extract the parameters of the model presented in \sref{opensyst}.

\subsection{Molecular dynamics}
\label{MD}

All our calculations have been performed for one monomer of the FMO complex of {\it C. tepidum} for which we have taken the X-ray structure 3BSD from the Protein Data Bank  as starting point.
This structure is embedded in a water box with equal amounts of sodium and chloride ions  added to obtain a 0.1 M/L salt concentration.
After energy minimization and equilibration, we have performed MD simulations of the ground state dynamics using the MD simulation package NAMD \cite{NAMD}. 
We performed simulations using either the AMBER99 \cite{ff99sb} or the CHARMM27 \cite{Charmm27_2,Charmm27} force fields.
For AMBER the force field of the BChls was adopted from \cite{ceccarelli:amber_forcefield} and was also used in Refs.~\cite{ShReVa12_649_,VaEiAs12_224103_}.
For CHARMM, we used the force field underlying Ref.~\cite{OlJaLi11_8609_} and has also been used recently in Ref.~\cite{GaShYe13_3488_}. 

For each force field, we simulated molecular dynamics at different temperatures and with various initial conditions, described in detail below. For each initial condition, we performed production runs of at least 300~ps duration, where the coordinates were saved every 5~fs. 
We describe further technical details of our MD simulations in \cite{supp:inf}. 

\subsubsection{Initial conditions for MD}
\label{MDinicond}
Here we briefly describe how we have generated different initial conditions (conformations) for the production run, in 
order to assess whether the usual assumption of ergodicity is justified in our context.
To this end we used three different equilibration sequences  (see \cite{supp:inf} for more technical details on the equilibration step).
\begin{itemize}
\item[A)] We equilibrated for 10~ns at temperature $T$.
\item[B)]  We equilibrated for 10~ns at 300~K, then set the temperature to $T$ and equilibrate there again for 10 ns.
\item[C)]  We equilibrated for 10~ns at 300~K, heat up to 310~K, then set the temperature to $T$ and equilibrate there again for 10 ns.
\end{itemize}

\noindent
For $T=$300 K only sequence A was used. 

\subsection{Extraction of the open system parameters}
\label{opensystparams}

The output of each MD simulation described in the previous section is a trajectory containing the positions of all atoms  in the FMO complex.

\subsubsection{Extraction of energy shifts}
\label{cdc_versus_zindo}

In a first step, we calculate so-called gap-energies of the BChls, the energy difference between ground and excited state along this trajectory. They are written as
\begin{equation}
\label{Etimetrace}
E_n(t)=\Eoffset+\delta_n(t),
\end{equation}
where $\Eoffset$ is an arbitrary energy offset common to all BChls and $\delta_n(t)$ are the deviations from this energy. 
From the $E_n(t)$ we then calculate mean site energies $\siteenergy_n$ and spectral densities $J^{(n)}(\omega)$ as detailed below. 
 We will focus here entirely on \emph{differences} between the gap-energies, for which calculation methods give reasonable results while absolute energies are problematic. Thus the offset $\Eoffset$ will drop out \footnote{Note further, that the offset $\Eoffset$ is important neither for excitation transfer nor for spectral densities (since here only deviations from the average enter).}.

We compare two methods to calculate the shifts $\delta_n(t)$:
\begin{enumerate}
\item[a)] The first approach is based on classical electrostatics. It has been introduced in \rref{AdMueMa08_197_} for a stationary situation and is denoted 'charge density coupling' (CDC).
In this approach, the energy shifts $\delta_n(t)$ of the transition energy caused by the field of $L$ point charges $Q_j$ in the surrounding dielectric medium are calculated as \cite{AdMueMa08_197_}
\begin{equation}
\delta_n=\frac{1}{\epsilon_{\rm eff}}\sum_{j=1}^L \sum_{i=1}^K \frac{\Delta q^{(n)}_i  \, Q_j}{|\vec{r}^{(n)}_i-\vec{R}_{j}|}.
\end{equation}
Here $\vec{R}_{j}$ is the position of the $j$th atom of the protein background, with charge $Q_j$, and $\vec{r}^{(n)}_i$ is the position of the $i$th atom of BChl $n$ with  transition charge $\Delta q^{(n)}_i= q^{(n)e}_i- q^{(n)g}_i$. The latter represents the change of charge density between the electronic ground- and excited state of the BChl, constructed as in \cite{MaAbRe06_17268_}.
The charges $Q_j$ of the protein background are contained as parameters in the description of the respective MD force field, and all positions $\vec{r}^{(n)}_i$, $\vec{R}_{j}$ are extracted from the MD trajectory.

The effective dielectric constant $\epsilon_{\rm eff}$  is used to describe the screening and local field effects by the dielectric environment.
 In accordance with Ref.~\cite{AdMueMa08_197_} we choose $\epsilon_{\rm eff}=2.5$. 
We note that $\epsilon_{\rm eff}$ can also be treated as a fit parameter, determined by comparison with experimental results.

\item[b)] 
In the second approach we use the semi-empirical {\bf Z}erner {\bf I}ntermediate {\bf N}eglect of {\bf D}ifferential {\bf O}verlap (ZINDO) method with parameters for spectroscopic properties (ZINDO/S) combined with the {\bf C}onfiguration {\bf I}nteraction {\bf S}ingles method (ZINDO/S-CIS).
For the calculation we used ORCA 2.9 \cite{ORCA}. 
The BChl molecule is treated quantum mechanical and the protein background within a sphere of 20 {\AA} radius around each BChl is taken into account via its point charges from the force field.
The influence of the radius of the sphere is discussed in some detail in \rref{OlKl10_12427_}.
To further speed up the calculations the phytyl tail of the BChls was replaced by a CH3 group. 
For the calculation of the gap-energy  the 10 highest occupied and the 10 lowest unoccupied orbitals are considered. 
This procedure is similar to the one used in Ref.~\cite{OlJaLi11_8609_}.

\end{enumerate}

\subsubsection{Extraction of the electronic Hamiltonian and static disorder:}
In a second step, we convert time-dependent energy shifts $\delta_n(t)$ thus obtained to 
the parameters that enter the open system model defined in  \sref{opensyst}.

The site energies of the BChls $\epsilon_n$ are simply taken as the mean values of the energy-gap trajectories, i.e.\
\begin{equation}
\label{eq:traj_mean}
\siteenergy_n= \overline{E_n(t)}=\Eoffset +\overline{\delta_n(t)},
\end{equation}
where the overbar denotes a time average over the full length of the MD trajectory. In a similar manner the excitation transfer elements $V_{nm}$ could be calculated from the MD trajectories by
$V_{nm}=\overline{V_{nm}(t)}$.
 In the present article, we concentrate on the calculation of the site energies and resulting spectral densities, deferring a discussion of these excitation transfer elements to a future publication.  

Besides the mean values, the trajectories $E_n(t)$ also provide a distribution for the probability of a certain energy.
Such distributions can be interpreted as static disorder \cite{KnScFi84_481_} and are presented in \sref{sec:trajects}. 

\subsubsection{Extraction of the spectral density:}
To determine a spectral density from the temporal fluctuations of energy shifts, 
we use the same formalism as in Ref.~\cite{VaEiAs12_224103_}. 
For BChl number $n$ a classical gap-correlation function
\begin{equation}
C_n^{\rm cl}(t,T)=\Big<\Delta_n(t+t_0)\Delta_n(t_0)\Big>_{t_0},
\label{classcorrelfct}
\end{equation}
is calculated, where $\langle \cdots \rangle_{t_0}$ denotes an average over $t_0$~\footnote{Numerically we use $C^{\rm cl}_n(t_k)=\frac{1}{M-k}\sum_{i=1}^{M-k}\Big(\Delta_n(t_i)-\overline{\Delta_n}\Big)\Big(\Delta_n(t_{i+k})-\overline{\Delta_n}\Big)$}.
We indicate explicitly that the classical correlation function \bref{classcorrelfct} is obtained from a MD simulation at a specific temperature $T$. 
Further
\begin{equation}
\Delta_n(t)=\delta_n(t) -\overline{\delta_n(t)} 
\end{equation}
is the energy gap fluctuation around the mean.

From this real correlation function the quantum spectral density is calculated via
\begin{equation}
J_n(\omega) = 2 \cdot  f(\omega,T)  \int_0^\infty dt\,  C_n^{\rm cl}(t,T) \cos (\omega t).
\label{specdenscalc}
\end{equation}
To avoid artefacts in the numerical Fourier transform caused by the finite length of our correlation function, we multiply the classical gap correlation function by a Gaussian window function. In our case this Gaussian was chosen such that it leads to a convolution of the actual SD with a Gaussian of standard deviation 20 cm$^{-1}$.

Since the spectral density entering our open system model is a temperature independent function, the prefactor $f(\omega,T)$ should ideally compensate any temperature dependence in \bref{classcorrelfct}. 

As has been theoretically shown in Ref.~\cite{VaEiAs12_224103_}, the choice 
$
f(\omega,T)= \frac{\hbar\omega}{2 k_B T }
$
is appropriate for our harmonic environment in the open system approach and
seems to be a reasonable choice in the case of the FMO complex.
More details on the complex question how to construct a quantum spectral density (or equivalently a quantum bath correlation function) are discussed in Ref.~\cite{VaEiAs12_224103_}. 

\section{Results}
\label{results}
\begin{table}[b]
\begin{tabular}{|l|cccc|}
\hline
method      &1a      & 1b  & 2a  &2b  \\
\hline
Forcefield   & CHARMM & CHARMM & AMBER & AMBER \\
Energy gap calc.   & CDC & ZINDO/S & CDC & ZINDO/S \\
\hline
\end{tabular}
\caption{Overview of calculation methods for molecular dynamics simulations and BChl energy gap calculations used in the present work.
\label{tab:methodcomparison}}
\end{table}
Table \ref{tab:methodcomparison} lists the different combinations of methods to perform molecular dynamics simulations and to extract excitonic Hamiltonians from the resulting data that we have compared here. The description of all results in this section refers to the method labelling in the table. 

In \sref{sec:trajects}  we first discuss general features of the site-energy trajectories resulting from all these methods. We then analyse the resulting mean site energies in \sref{site_energies}, correlations of energies in \sref{correlations} and spectral densities in \sref{specdens}. 

\subsection{General features of the trajectories}
\label{sec:trajects}
\begin{figure*}[tbp]
\includegraphics[width=18cm]{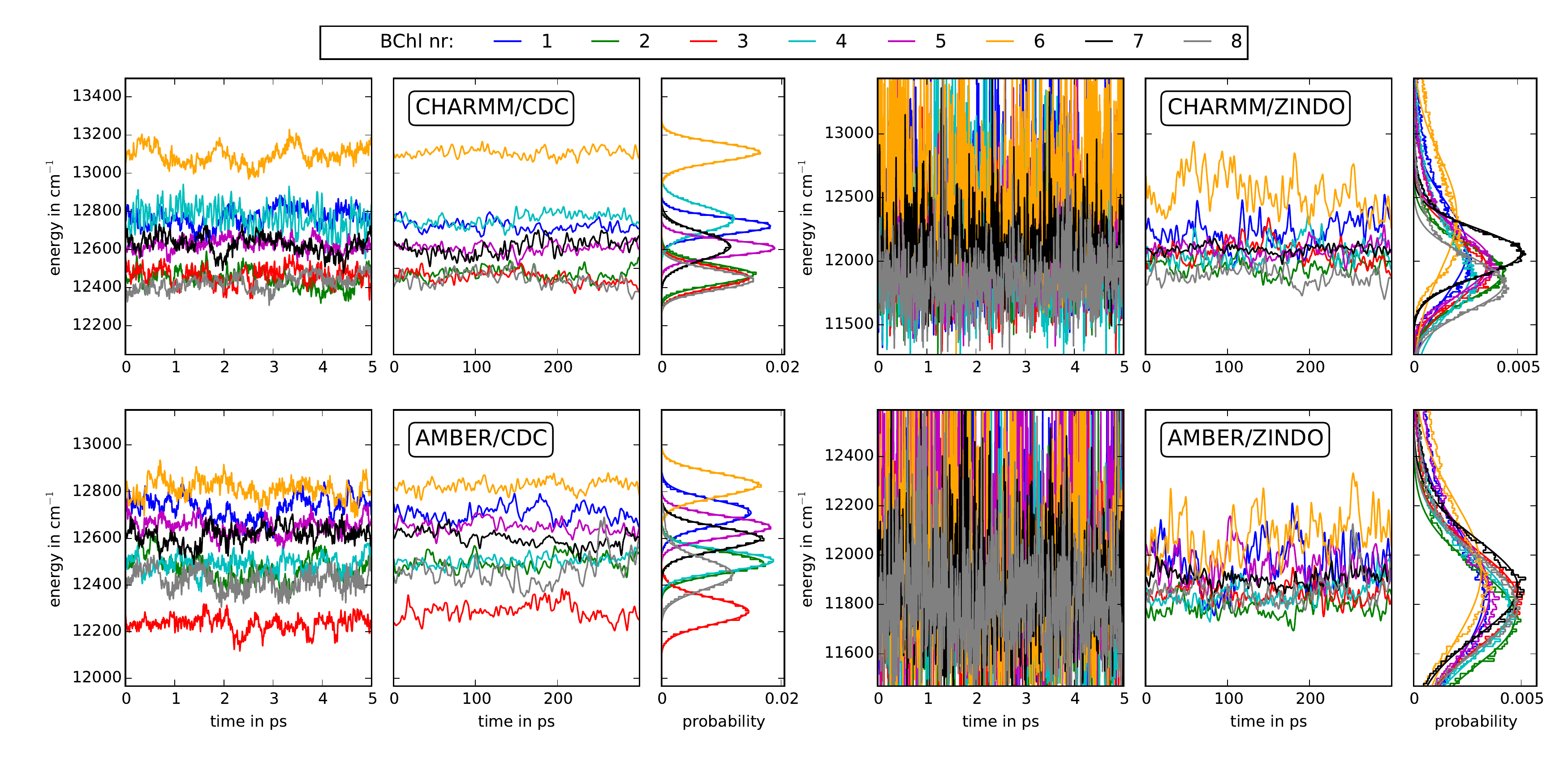}
 \caption{\label{fig:energy_timeseries_300K}
Example of the time-dependent site energies at temperature 300K. 
In each quadrant: (left panel) a zoom onto a 5ps segment. (middle panel) Full 300 ps trajectory smoothed over windows of $5\ \rm{ps}$ \cite{footnote:smoothing}, which is the time interval shown in the left panel. (right panel) Probability to find a given gap energy within an interval of width $10\ \rm{cm}^{-1}$. 
The assignment of colors to pigment sites is shown in the legend and followed throughout this article.
}
\end{figure*}

Before we discuss the parameters for the open system model it is worthwile to take a look at the trajectories obtained with the different approaches. 
They are shown in \fref{fig:energy_timeseries_300K} for all four methods at temperature 300K.
Trajectories for the other BChls, at further temperatures and using different initial conditions can be found in the supporting information \cite{supp:inf}. 
The trajectories presented in  Fig.~\ref{fig:energy_timeseries_300K} already allow us to discuss some general features.

For each method the left panel contains a 5 ps segment of a trajectory,
including all sampled data points at 5 fs intervals. They illustrate the fluctuations of the gap-energy on these short time scales.
When comparing the CDC method (left column) with the ZINDO calculations  (right column) it becomes evident that the fluctuations in the CDC method are roughly an order of magnitude smaller than those obtained with ZINDO.
We will come back to this point several times in this work.
Note that the absolute energy in the CDC method can be shifted arbitrarily, since it only delivers gap-energy differences, see \sref{cdc_versus_zindo}.
To provide a feeling for the slow long-time dynamics, we show in the middle panel for each method a 300 ps trajectory that was smoothed over windows of $5\ \rm{ps}$~\cite{footnote:smoothing}.
The magnitude of the fluctuations in these curves is much smaller than that in the left panels.
However, for the ZINDO it is again  larger than for CDC.
Finally, we show the distribution to have a certain gap-energy in the trajectory in the right panels, with a Gaussian fit as a guide to the eye. As expected from the left and middle panel, the width of the distributions for the ZINDO calculations is roughly an order of magnitude larger than that of the CDC calculations.

We find that differences between these general features of the trajectories for different force fields (upper and lower row in Fig.~\ref{fig:energy_timeseries_300K}) are less pronounced than for  different methods to calculate the gap-energies (left and right column).
Note that MD simulations with different force fields employ different initial states and different trajectories.
This apparently causes the ordering of the peaks in site-energy distributions of the CDC method (right panels, left column) to differ between the force fields.
However, although we have used \emph{the same} MD trajectory within each forcefield (i.e.~within each row), the ordering of the peaks also differs between CDC and ZINDO.
 This comparison is made difficult by the large widths found with ZINDO, but is still possible e.g.~for site $2$.
We will consider this point in more detail when discussing the site energies in \sref{site_energies}.

\subsubsection{Interpretation}
\label{trajectory_interpretation}
\begin{figure}
\includegraphics[width=8cm]{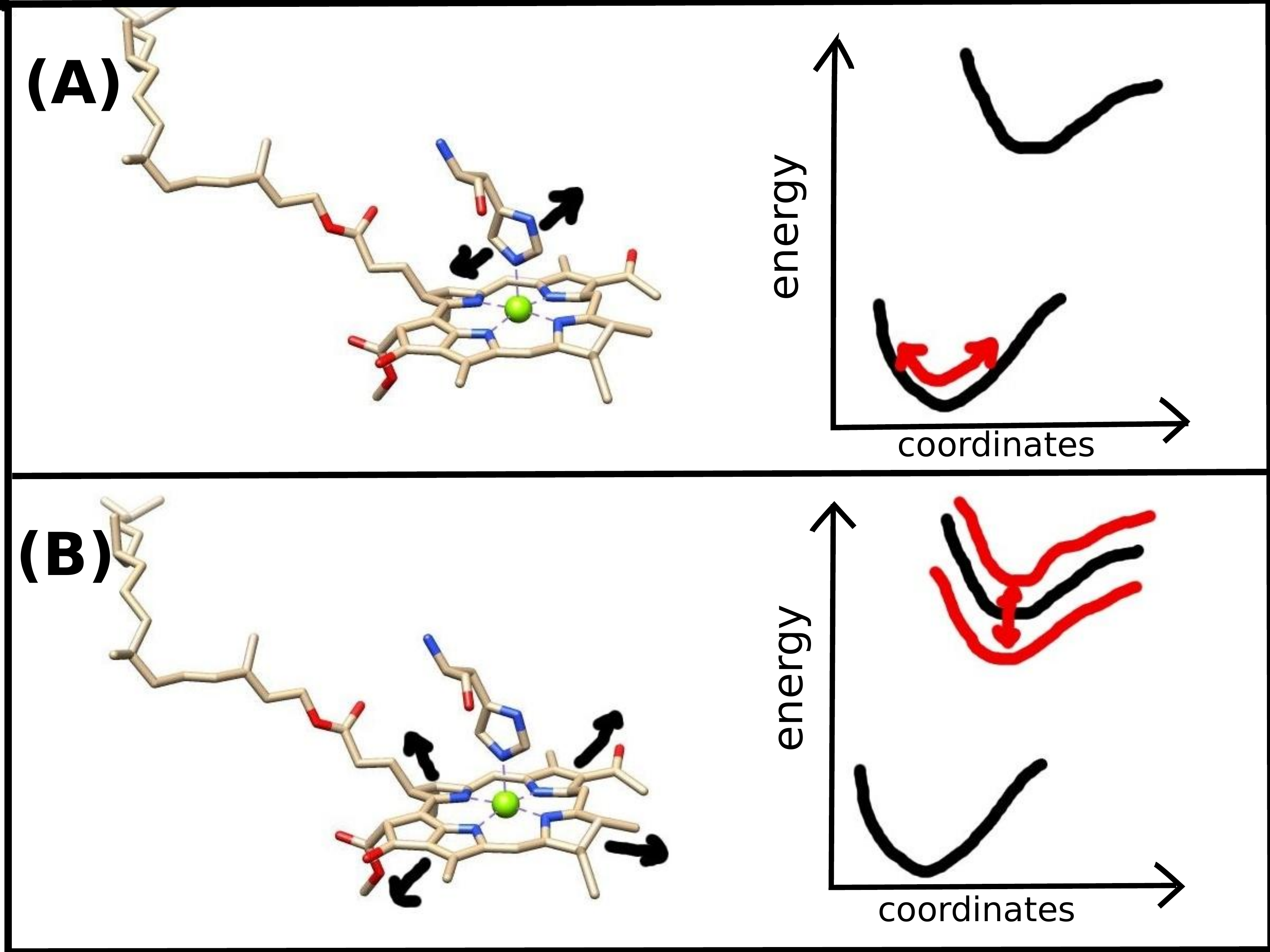}
\caption{\label{fig:sketch_internal_vs_external}  We sketch how nuclear dynamics internal- or external to the BChl differently affect its gap-energy.
(A) Internal vibrations. If the molecule is deformed the energy gap changes according to the difference of the ground and excited state Born-Oppenheimer surfaces. (B) External vibrations: The motion of the protein leads to changes in the electrostatic interaction, which basically results in a energy shift of the difference between the two BO surfaces. In CDC only the latter effect is taken into account.
Note, that protein dynamics can also induce dynamics within the BChls.  }
\end{figure}

One can understand the difference in magnitude between the CDC and the ZINDO calculations, when considering the particular molecular properties that each method takes into account. In the ZINDO method, one calculates the energy difference between two Born-Oppenheimer surfaces. 
Here {\it internal} nuclear coordinates of the BChls play an important role.
Small changes away from the equilibrium position can lead to large changes in the gap-energy.
This is sketched in \frefp{fig:sketch_internal_vs_external}{a}, for the case of only one nuclear coordinate and shifted harmonic potentials.
On the other hand nuclear coordinates {\it external} to the BChl affect the CDC only in very weak fashion. 
They only slighty modify the distances between the partial charges of the BChls and the protein.  
This is sketched in \frefp{fig:sketch_internal_vs_external}{b}.

\subsubsection{Lower temperatures}
It is instructive to repeat the calculations above for lower temperatures.
We present results for 77K in Fig.~\ref{fig:energy_timeseries_77K},
and results for 200K in the supplemental information \cite{supp:inf}.
\begin{figure*}[tbp]
\includegraphics[width=18cm]{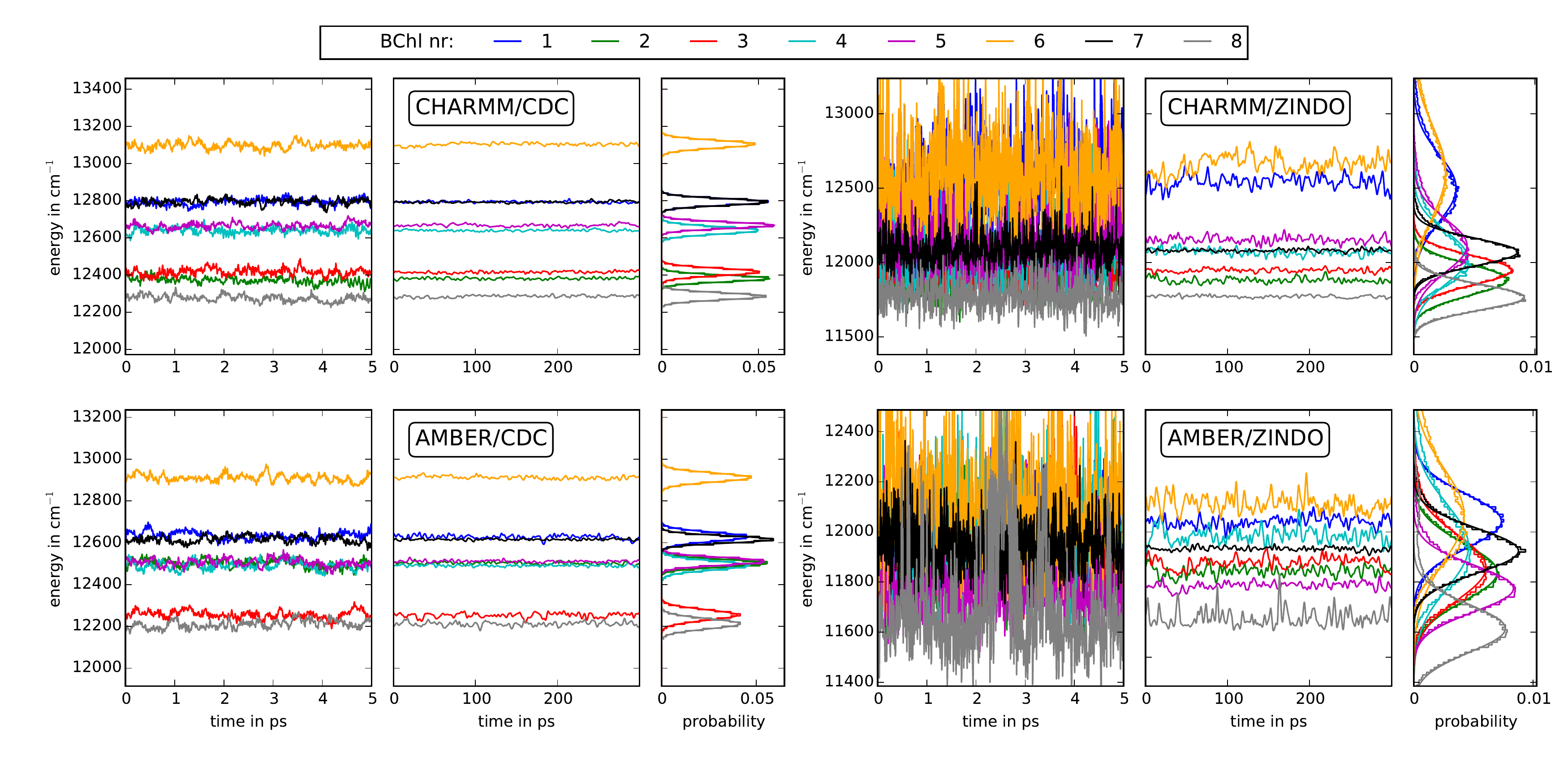}
 \caption{\label{fig:energy_timeseries_77K} 
Same as Fig.~\ref{fig:energy_timeseries_300K}) but for 77K.
}
\end{figure*}
The most striking new feature, is that the magnitude of gap-energy fluctuations, and thus the width of their distributions, is narrower than for the 300K trajectories.
We find a roughly linear relationship of this width with the temperature $T$, for our three values of $T$.

Similar to the 300K case, the energetic ordering of the different sites still varies between all methods for $T=77$ K, and is further different from the ordering at 300K.
In contrast to room-temperature, we can clearly assign a relative order to the peaks of the energy-gap distributions also for the ZINDO calculations. 
Even when just varying the initial conditions as described in \sref{MDinicond}, we find variations in the peak ordering.

While the width of the gap-energy distributions for all monomers are roughly of the same size at $T=300$ K, at $T=77$ K some widths are nearly a factor two broader than others for ZINDO calculations. This holds in particular for site 6. We believe this is linked to the energy drift visible for site 6 in the middle panel of the upper right quadrant in Fig.~\ref{fig:energy_timeseries_77K}, at the beginning of the trajectory, and attribute it to conformation changes of the BChl, which cause a large change of the internal BChl energy-gap. Note that the CDC calculations in Fig.~\ref{fig:energy_timeseries_77K} do not show the drift, although they are based on the same trajectory, presumably since CDC is less sensitive to internal BChl coordinates as argued in the previous section.

Let us remark that Ref.~\cite{Aghtar:glyzerol}, where a ZINDO approach (with CHARMM) has been used, also finds the site energy width of some BChl to be much broader than of others. A trimeric complex has been studied in that work, which further finds that the energy width (and mean) of a certain BChl (e.g.\ no.~3) also changes drastically between the three monomers.

\subsection{Site energies }
\label{site_energies}

\begin{figure}[tbp]
\includegraphics[width=0.95\columnwidth]{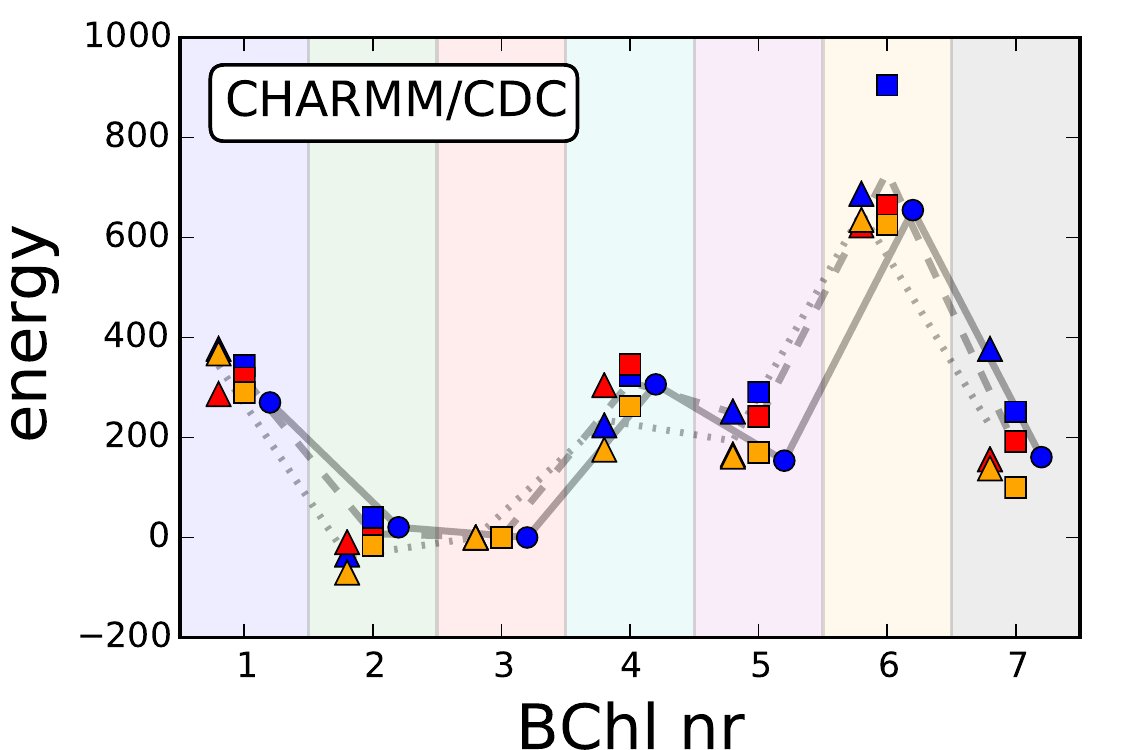}
\includegraphics[width=0.95\columnwidth]{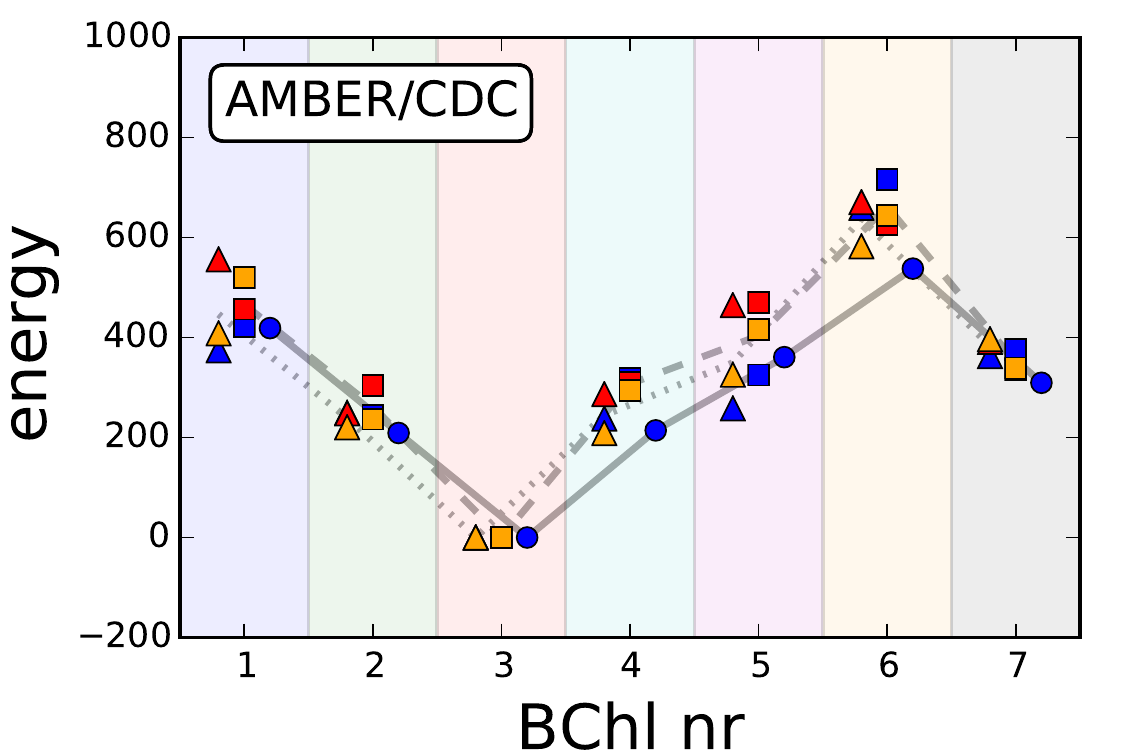}
\caption{  Mean site energies obtained from CDC simulations at 77K (triangles), 200K (squares) and 300K (circles) for the two different force fields. 
Upper panel: method 1a, lower panel: method 2a.
The colors of symbols discriminate the different initial conditions (blue: A, red: B, orange: C).
The color shading of the background identifies different sites as in \fref{fig:energy_timeseries_300K}.
 \label{site_energies_figure} 
  }
\end{figure}

From the trajectories presented in \fref{fig:energy_timeseries_300K}-\ref{fig:energy_timeseries_77K} above, we now construct the mean transition energies of the BChls according to Eq.~\eqref{eq:traj_mean}.
The results are shown in Fig.~\ref{site_energies_figure}, for CDC trajectories at 77 K, 200 K and 300 K, including variations for other initial conditions. 
We show the corresponding ZINDO results only in the supporting information \cite{supp:inf}, since the standard deviation of the mean transition energies there exceeds average energy differences between the sites, complicating an interpretation.
 
We can see in \fref{site_energies_figure} that all methods roughly agree on the overall site energy trend. The most striking difference between CHARMM (method 1a, upper panel) and AMBER (method 2a,  lower panel), 
is that sites 2 and 5 have significantly lower energies using the former forcefield. Another important observation is that also different initial conditions cause significant differences in mean energies.
Since the underlying energy distributions are narrower than the differences between these means, as shown in \fref{fig:energy_timeseries_77K}, 
this indicates that the different equilibration procedures that define our initial conditions (see \sref{MDinicond}) cause the system to end up in different {\it local} energy minima.
We do not exclude that this effect also occurs at 300K, but have not sampled different initial conditions there.  

Finally, we would like to point out that for some initial conditions, neither site three nor four has the lowest gap-energy. These simulations then do not meet expectations arising from the physiological function of the FMO, 
as transporting energy towards the reaction centre near sites three and four. However, most reported MD studies agree on site three having the lowest energy, as do most of our simulations. 

\subsubsection{Comparison with previous results}

The literature values of the site energies of the various BChls show a large spread. 
Our results fall within this spread.
Here we only discuss briefly the results of another CDC calcuation performed on the same species $\it C. tepidum$ (but a different PDB structure) using the CHARMM forcefield \cite{RiMoMa13_5510_}.
Their values are quite similar to our AMBER/CDC results at 300K, the largest difference being at site 6 where our value is roughly a factor 1.8 larger.

Our AMBER/CDC results also are close to the values of Ref.~\cite{AdRe06_2778_} with the notable difference that their BChl 1 has a lower energy than BChl 2.
 
\subsection{Correlations of site energy fluctuations}
\label{correlations}
In the definition of our open quantum system model in \sref{opensyst} we have assumed independent harmonic baths for the different BChls.
Although not necessary, this assumption makes calculations easier and is often used.
Previous investigations indicate that indeed correlations between the gap-energy fluctuations at different BChls are very small \cite{OlStSc11_758_}. 

\begin{figure}
\includegraphics[width=1.0\columnwidth]{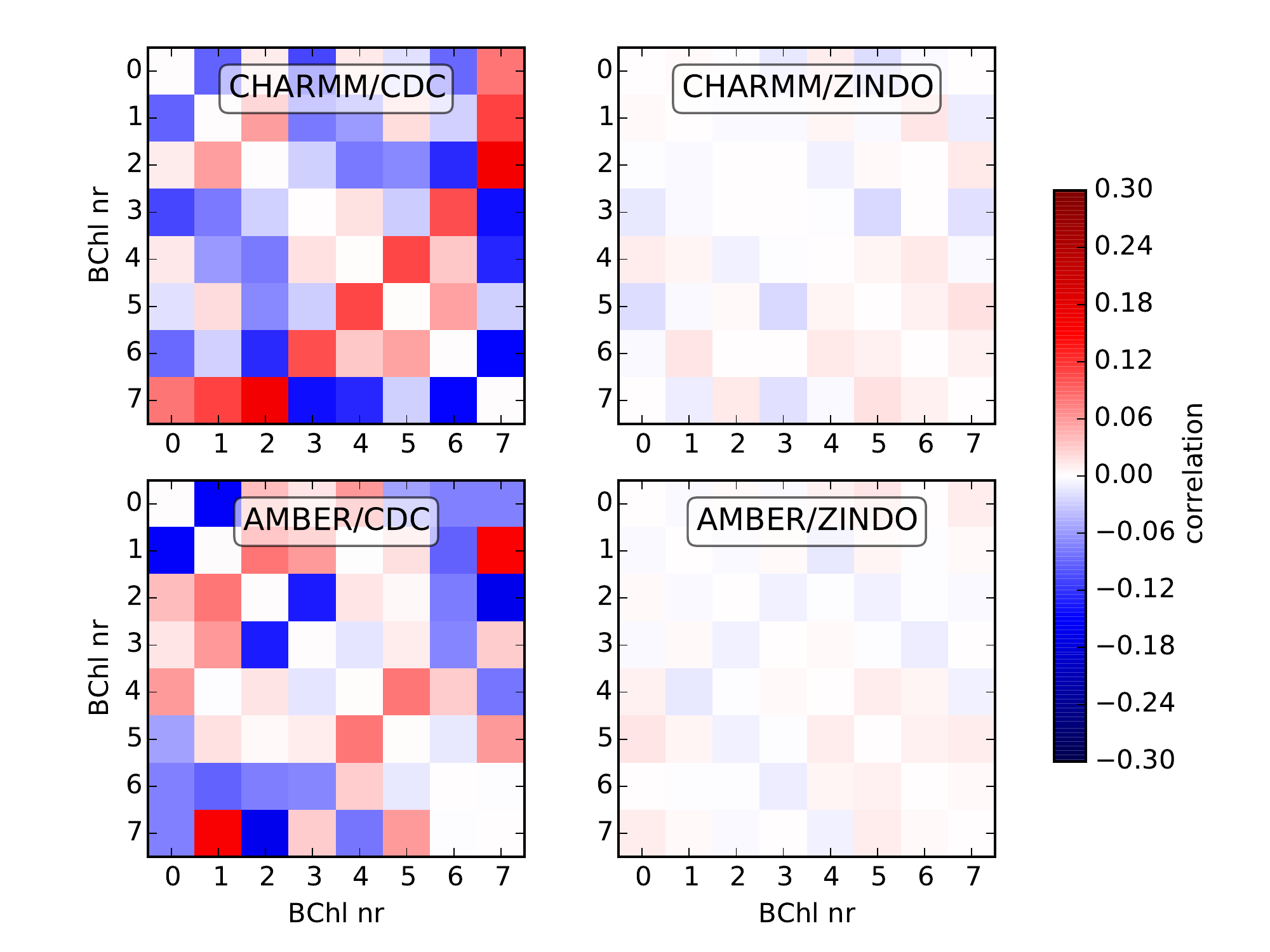}
\caption{Correlations $C_{nm}$ according to \eref{eq:C_ij} between the energy gap fluctuations at different sites $n$, $m$, using $T=300K$. The diagonals fulfill $C_{nn}=1$ and are set to zero.  
\label{fig:correlations_300K}
} 
\end{figure}

\begin{figure}
\includegraphics[width=1.0\columnwidth]{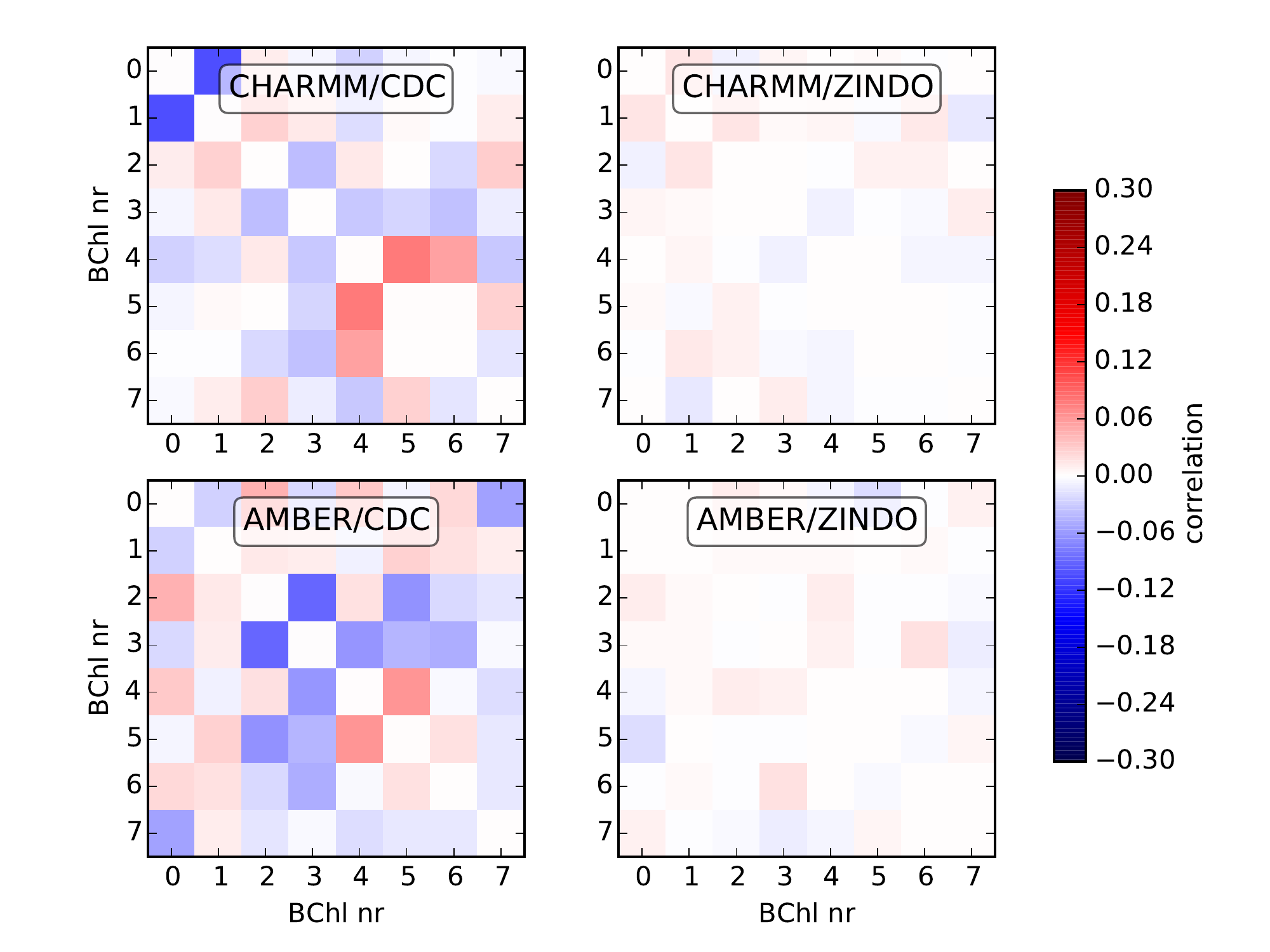}
\caption{Same as Fig.~\ref{fig:correlations_300K} but for 77K.
\label{fig:correlations_77K}
} 
\end{figure}

In the following we investigate how strong the gap-energy fluctuations at different sites are correlated using all methods in \tref{tab:methodcomparison}.
To this end we evaluate the correlation function between sites $n$ and $m$ 
\begin{equation}
\label{eq:C_ij}
C_{nm}=\frac{\overline{\Delta_{n}(t)\cdot\Delta_{m}(t)}}{\sigma_n\cdot \sigma_m},
\end{equation}
where $\sigma_n=\big(\,\overline{(\Delta_n(t))^2}\,\big)^{1/2}$ is the standard deviation of the gap-energy distribution of BChl $n$, associated with the trajectory. 
Figs.~\ref{fig:correlations_300K} and \ref{fig:correlations_77K} show the resulting correlations for 300K and 77K, respectively.

Clearly, the correlations in the CDC method (left column) are much larger than those obtained with the ZINDO calculations.
Decreasing the temperature also decreases the correlations, an effect that is
more pronounced for the CDC method. 
To quantify these observations, we provide the maximal correlation, the sum of the correlations for all methods, temperatures and initial conditions in the supporting information \cite{supp:inf}.

 The fact that the correlations for CDC are much stronger than for ZINDO is consistent with our observations in \sref{trajectory_interpretation}, that
 ZINDO fully samples internal potential energy surfaces, while they only have a weakened effect in CDC. Hence
the main fluctuations in CDC are due to protein motion. Since it is fair to assume that protein motion can affect several BChl in a correlated fashion, while internal vibrations of separate BChls are uncorrelated,  fluctuations that are more strongly affected by internal vibrations (ZINDO) are less correlated between sites
 \footnote{To see this clearly, we write the fluctuations in ZINDO as an internal part $\Delta_n^{\rm BChl}(t)$ and an external part $\Delta_n^{\rm Prot}(t)$, thus $\Delta(t)=\Delta_n^{\rm BChl}(t)+\Delta_n^{\rm Prot}(t)$.
Then $\sigma_n^2=\overline{\Delta_n^{\rm BChl}(t)^2}+\overline{\Delta_n^{\rm Prot}(t)^2}+2\overline{\Delta_n^{\rm BChl}(t)\Delta_n^{\rm Prot}(t)} $.
We expect that internal and external induced energy shifts are not strongly correlated and therefore use $\overline{\Delta_n^{\rm BChl}(t)\Delta_n^{\rm Prot}(t)}\approx 0$. 
As discussed when presenting our energy-gap trajectories in section \ref{sec:trajects}, we also have $\overline{\Delta_n^{\rm BChl}(t)^2} \gg \overline{\Delta_n^{\rm Prot}(t)^2}$.
Thus $\sigma_n\approx \sigma_n^{\rm BChl}$, and the denominator in Eq.~(\ref{eq:C_ij}) becomes $\sigma_n^{\rm BChl}\sigma_m^{\rm BChl}$.
In the numerator we expect that internal vibrations of different BChls are uncorrelated and that they are also uncorrelated with Protein-dynamics. Then we find $\overline{\Delta_n(t)\Delta_m(t)}\approx \overline{\Delta_n^{\rm Prot}(t)\Delta_m^{\rm Prot}(t)}$ and finally 
 $C_{nm}\approx\frac{\overline{\Delta_n^{\rm Prot}(t)\Delta_m^{\rm Prot}(t)}}{\sqrt{\overline{\Delta_n^{\rm BChl}(t)^2}}\sqrt{\overline{\Delta_m^{\rm BChl}(t)^2}}}$.
This is small as one sees using similar aguments as above. 
 For CDC, assuming $\Delta_n^{\rm BChl}(t)\approx \Delta_n^{\rm Prot}(t)$ one has $\Delta_n^{\rm BChl}(t)^2\approx\Delta_n^{\rm Prot}(t)^2$ in the denominator of that expression. Thus $C_{nm}$ will be larger as for the ZINDO case.
}.

\subsubsection{Comparision with previous results}
\label{comparison_with_others}

Our results here are consistent with a previous study on the FMO Chlorobaculum tepidum (PDB code 3ENI)~\cite{OlStSc11_758_}, 
which also essentially indicate the absence of correlations using ZINDO, with $C_{nm}\lesssim0.05$ throughout. 
However, here we also see that correlations exist, if one is only interested in the \emph{external} influence, that can be separated using CDC.

\subsection{Spectral densities}
\label{specdens}
\begin{figure*}
\includegraphics[height=20cm]{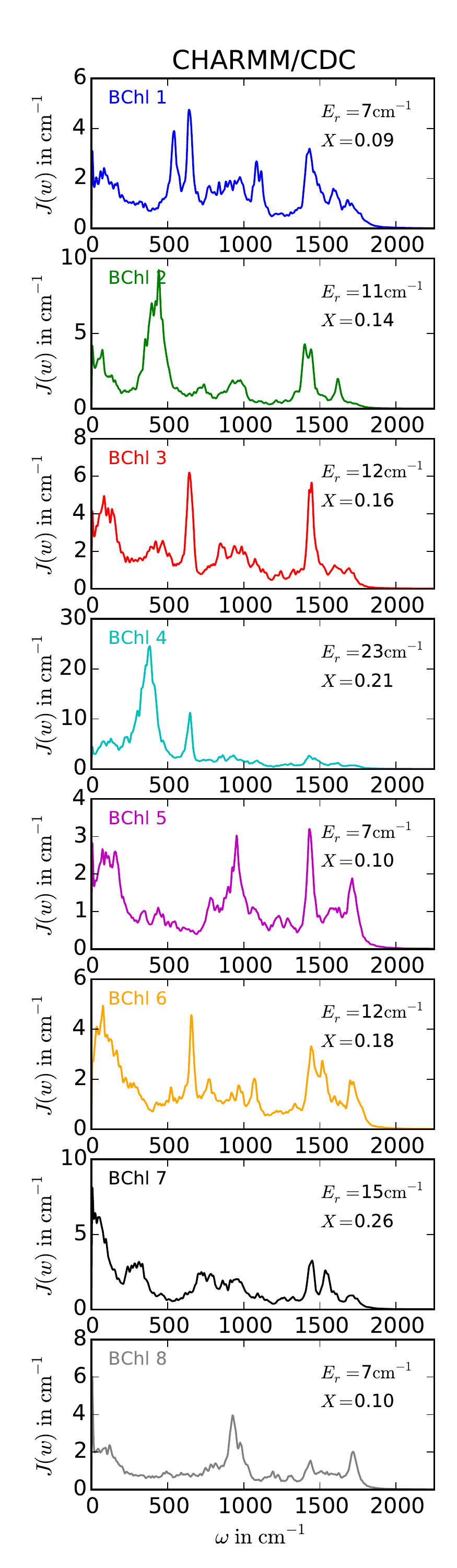}
\includegraphics[height=20cm]{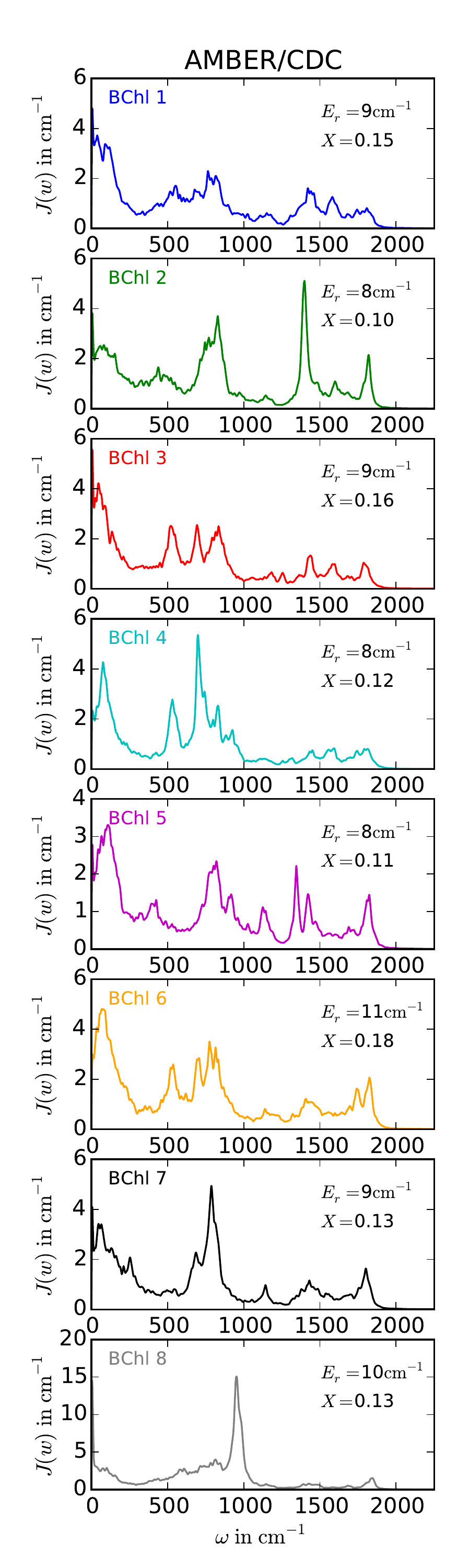}
\caption{\label{fig:SD_1a_2a_all_sites_300K} Spectral densities at 300K for method 1a (left) and method 2a (right), separately for each site.
We also provide the values of the reorganization energy $E_r$, \eref{reorg_energy}, and the integrated 'Huang-Rhys-factor' $X$, \eref{huang_rhys}.  }
\end{figure*}
\begin{figure*}
\includegraphics[height=20cm]{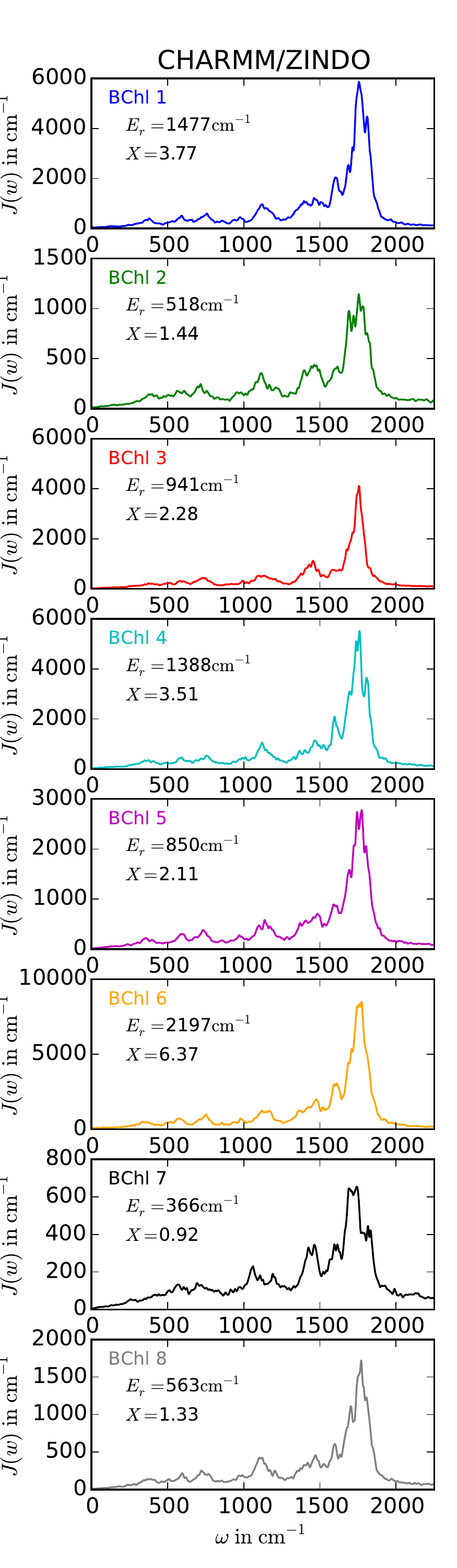}
\includegraphics[height=20cm]{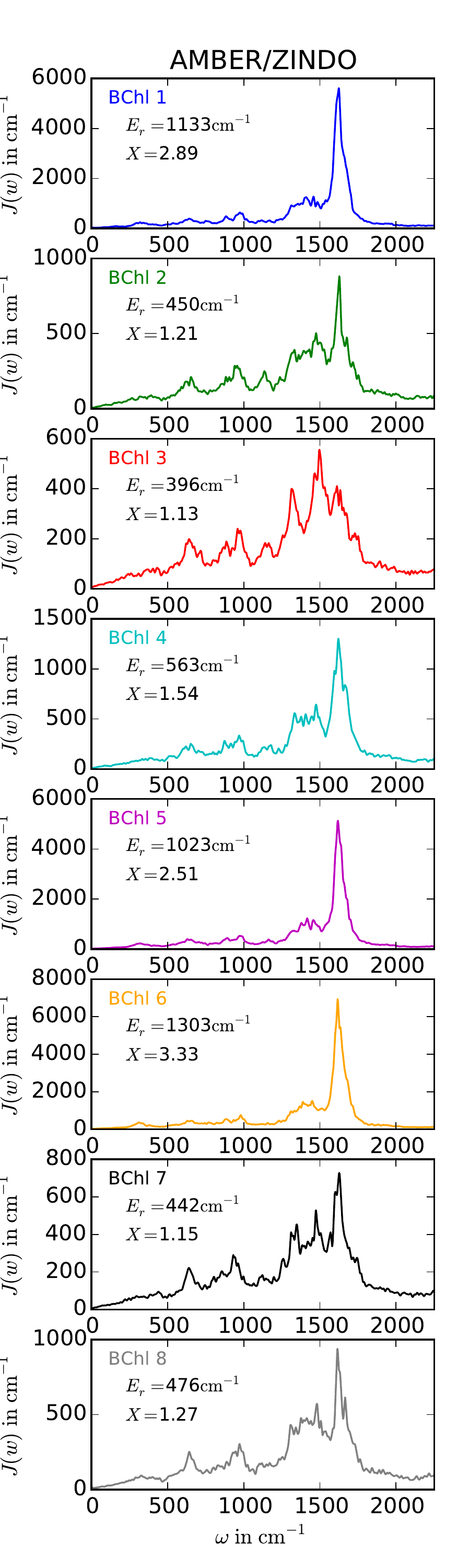}
\caption{\label{fig:SD_1b_2b_all_sites_300K} Same as \fref{fig:SD_1a_2a_all_sites_300K}, but using method 1b (left) and method 2b (right).}
\end{figure*}
After our detailed discussion of features in gap-energy trajectories, such as the width of fluctuation distributions and the absence of inter-site correlations, we now proceed to the extraction of spectral densities

We determine spectral densities from trajectories of energy-gap fluctuations according to \eref{classcorrelfct} and \eref{specdenscalc}~\footnote{Convergence regarding the time-average in \eref{classcorrelfct} was confirmed by comparing results obtained from the full $300$ ps trajectories with those from 
averaging over randomly chosen $5$ps sub-segments, finding identical results.}.

The spectral densities of all 8 Bchl molecules calculated with CDC (methods 1a and 2a) at 300K  are shown in Fig.~\ref{fig:SD_1a_2a_all_sites_300K}.
Those  calculated with ZINDO  (methods 1b and 2b) at 300K  are shown in Fig.~\ref{fig:SD_1b_2b_all_sites_300K}.
In each panel we also provide integral measures that are often used to assess the 'total strength' for the interaction of the respective BChl with the bath.
The first measure is the so-called reorganization energy, defined as 
\begin{equation}
E_r = \frac{1}{\pi}\int_c^{\infty} d\omega\, J(\omega) / \omega,
\label{reorg_energy}
\end{equation}
the second one is the total Huang-Rhys factor
\begin{equation}
X = \frac{1}{\pi}\int_c^{\infty} d\omega\, J(\omega) / \omega^2.
\label{huang_rhys}
\end{equation}
We choose a small finite lower bound of $c=20 \rm{cm}^{-1}$ for the frequency integral, to avoid the divergence at $\omega=0$
\footnote{The choice of $c$ strongly affects $X$ and $E_r$. We choose it as the lowest value for which, \emph{increasing} $c$ does not significantly alter $X$. 
Note also that the correct description of frequencies near zero would require much longer trajectories, such that ignoring these frequencies is more physical.
}.
Here we exemplarily discuss the SDs obtained for 300K, the SDs for other temperatures and varied initial conditions are shown in the supporting information \cite{supp:inf}.

Before we separately discuss detailed features of SDs obtained from CDC and ZINDO, we highlight the pronounced differences between the CDC and the ZINDO calculations, i.e.\ between Fig.~\ref{fig:SD_1a_2a_all_sites_300K} and Fig.~\ref{fig:SD_1b_2b_all_sites_300K}. 
Two facts immediately catch the eye:
Firstly, the height of the peaks grows with increasing energy in the ZINDO calculation, while it roughly remains the same or even decreases in CDC.
Secondly, the overall magnitude of SDs obtained in CDC calculations is much smaller than of those obtained using ZINDO.
The latter, is also reflected in reorganisation energies and total Huang-Rhys factors.
This is again due to a distinction of internal vibrational modes of the BChls and external motion of the protein.
As discussed in \sref{trajectory_interpretation}, internal vibrations contribute to the SDs in CDC only indirectly via interactions with the protein, in contrast to ZINDO.
Since in both cases motion is determined by the same (ground state) energy surface, vibrational \emph{frequencies } of internal modes do however still affect the CDC results. 
Therefore we expect to see peaks in CDC spectra at similar positions as for ZINDO, but with much smaller magnitude. 
A clearly visible example is the peak near $1500$ cm$^{-1}$, present in both CDC and ZINDO.
 
\subsubsection{Spectral densities from CDC}
In general the SDs that we have obtained with CDC only show minor variations between different temperatures and different initial conditions.
We emphazise this point, since for the ZINDO calculations we will see large variations in the magnitude.

The results of \fref{fig:SD_1a_2a_all_sites_300K} can be interpreted as capturing the coupling to a bath that is \emph{external} to the BChl, based on our discussion above.
This has been more rigorously shown by Jing {\it et al.}~\cite{JiZhLi12_1164_}, who applied CDC and ZINDO/S to MD simulations of the reaction center of purple bacteria. They compared ZINDO calculations with and without partial charges on the surrounding protein, and found that spectral densities based just on the \emph{difference} between these two fluctuations agree with those obtained from CDC.

\subsubsection{Spectral densities from ZINDO}
\label{specdens_ZINDO}
\begin{figure}
\includegraphics[width=9cm]{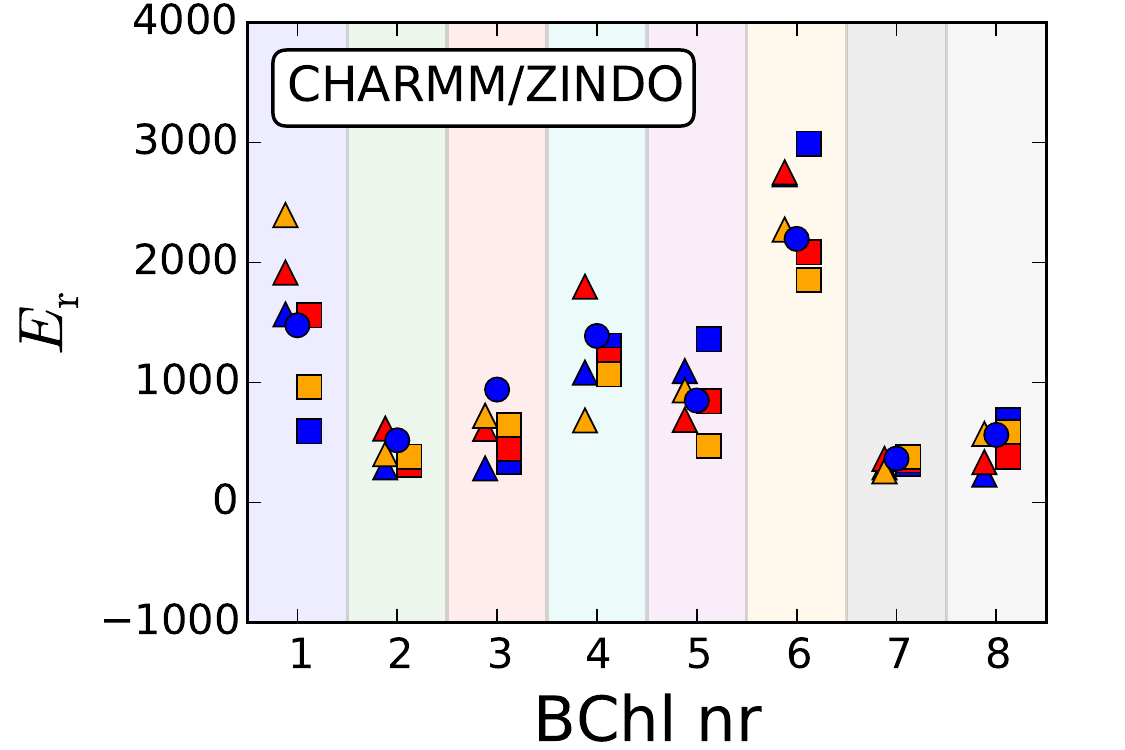}
\includegraphics[width=9cm]{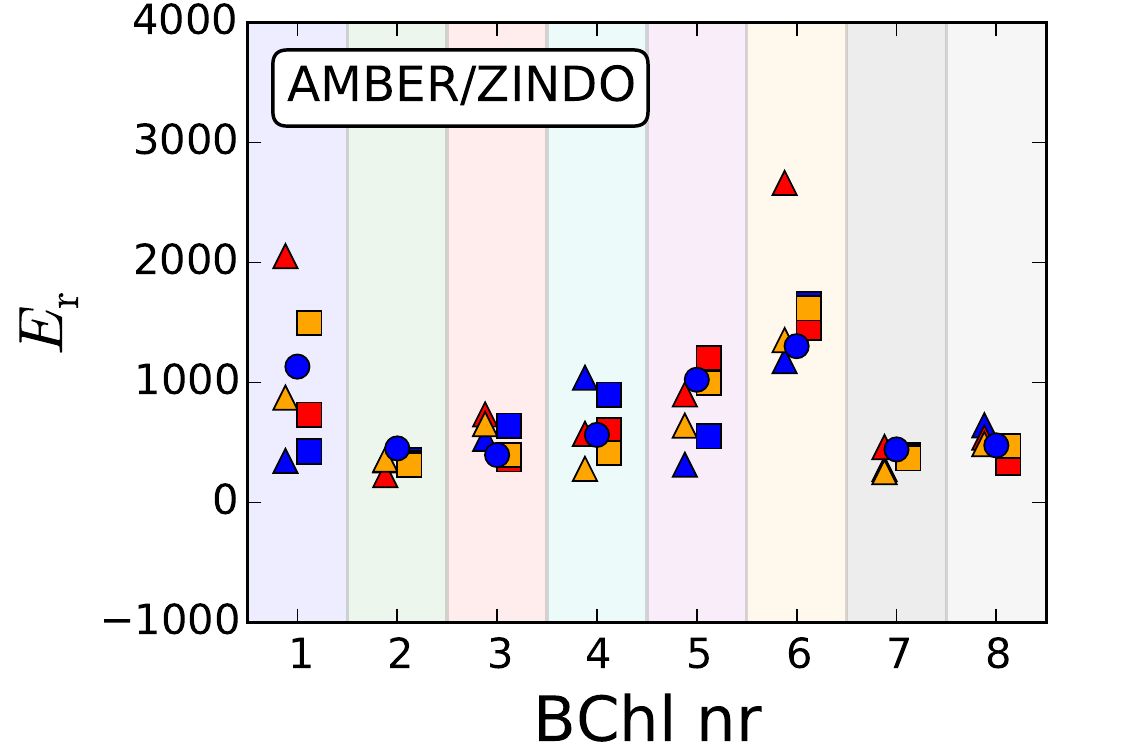}
\caption{\label{fig:SD_Reorgs_all}Reorganisation energies, \eref{reorg_energy}, for all spectral densities calculated using the ZINDO method. Top: method 1b, bottom: method 2b, symbols as in \fref{site_energies_figure}.
}
\end{figure}

The overall shape for SDs of all BChls is very similar, independent of temperature and initial conditions. 
However, we notice two variations: 1) The highest energy peak is sometimes strongly suppressed, for example in method 2b at 300K for site site three and four. 
2) The magnitude of all SDs varies considerably from site to site as can be seen in Fig.~\ref{fig:SD_1b_2b_all_sites_300K} for 300K, and in the supplemental information \cite{supp:inf} for 77K and 200K.
The smallest and largest SD differ by more than an order of magnitude (as is also evident in the reorganization energy and the total Huang-Rhys factor).
For a more quantitative picture, Fig.~\ref{fig:SD_Reorgs_all} shows reorganization energies for all trajectories analysed \footnote{We obtain the same qualitative picture when plotting the maximal peak height or the total Huang-Rhys-factor.}.
One sees large variations from site to site, within each temperature and for the different initial conditions. 
Nonetheless, as we discuss in the next section, the average of these results (per site) are roughly consistent with other theoretical studies and available experiments.
 The variations thus highlight that great care has to be taken in the detailed implementation of the schemes employed here. Our interpretation is as follows: The structure relaxation of the MD simulations yields a variety of BChl conformations with significant probability. The shift between ground- and excited energy surfaces may depend on the conformation, yielding different amplitudes in spectral densities. Possibly more important, BChl conformations far from the equilibrium configuration of an isolated BChl might occur, for which the employed ZINDO method could give erroneous results. 
This may explain outliers in \fref{fig:SD_Reorgs_all}, as also conjectured in \cite{JiZhLi12_1164_}. More clarity could be gained by a joint analysis of reorganisation energies and BChl geometries, which is beyond the scope of the present work.
Another possible cause of the spread observed in \fref{fig:SD_Reorgs_all} can be water molecules that  enter the protein and reach the vicinity of BChls. 
Since we did not save the positions of water molecules in our simulations, we defer an investigation of this possibility to future work.

Finally, let us point out that there is a pronounced shift between the high energy peak for the different force fields.
For the CHARMM/ZINDO results this peak is located at an energy approximateley 200 cm$^{-1}$ higher than for the AMBER/ZINDO.
This is consitent with the results found in the previous literature.

\subsubsection{Comparison to  other theoretical calculations}

\begin{figure}
\includegraphics[width=\columnwidth]{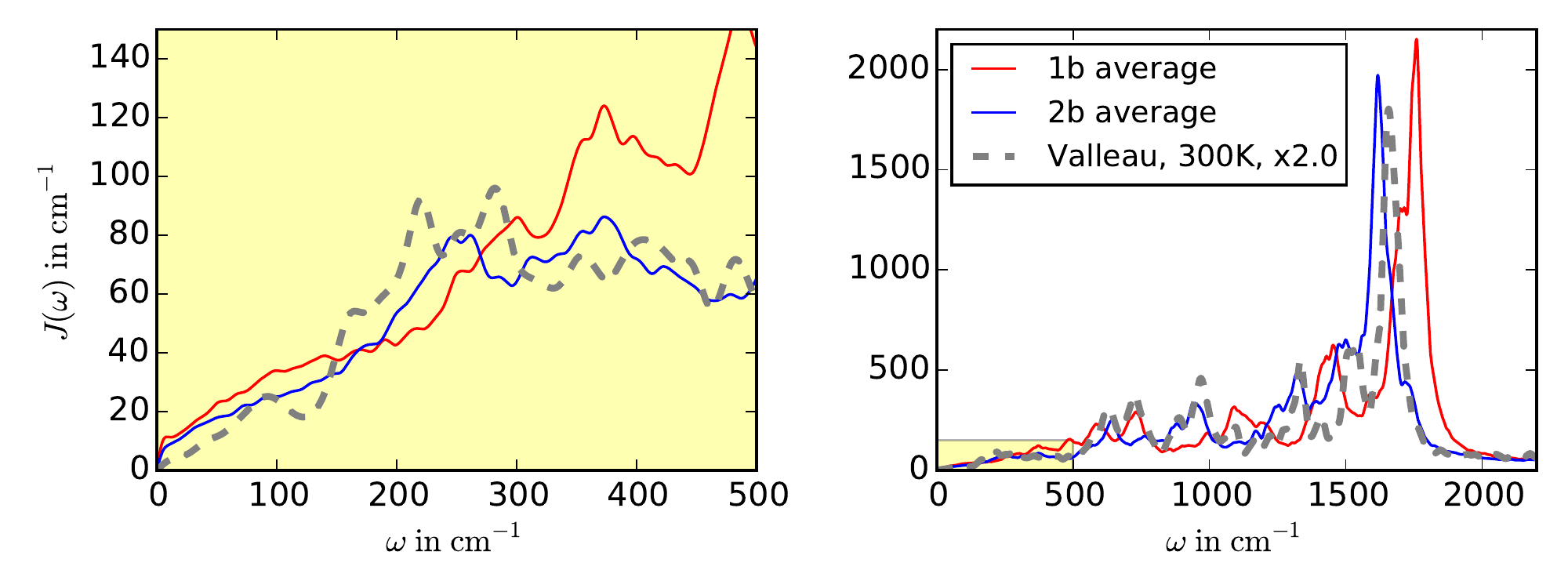}
\caption{\label{fig:Comparison_SD_with_theory_site3}Spectral densities of site 3, comparison of our results and theory from \cite{VaEiAs12_224103_}. Left: low frequency part. Right: full spectral density. The red and blue curves are an average of all SDs that we have calculated for method 1b and 2b, respectively. The gray, dashed curve is the result of  \cite{VaEiAs12_224103_} multiplied by a factor of two.
}
\end{figure}

Let us first compare our results with theoretical studies in the literature, which employed only a single of the methods compared here.
We begin with Valleau {\it et al.} \cite{VaEiAs12_224103_}, who present spectral densities that agree quite well with experiment and are independent of temperature. For a comparison, one however has to keep in mind that Ref.~\cite{VaEiAs12_224103_} models the slightly different structure of {\it Prosthecochloris aestuarii} (PDB ID: 3EOJ) \cite{ShReVa12_649_}, containing only 7 BChl.
Our results of methods 1b and 2b are shown in Fig.~\ref{fig:Comparison_SD_with_theory_site3} together with those of Ref.~\cite{VaEiAs12_224103_} for monomer 3.
We have averaged all SDs obtained from different temperatures and initial conditions.
The resulting SDs are about a factor of two larger than reported in~\cite{VaEiAs12_224103_}.
We see that the results of method 2b are more similar to~\cite{VaEiAs12_224103_}, as expected since method 2b and \rref{VaEiAs12_224103_} both rely on the same force field (AMBER).
Also the width of the energy-gap distributions (see \cite{supp:inf}) is comparable. 

Another MD study of {\it Prosthecochloris aestuarii} was reported in \cite{OlJaLi11_8609_}, using CHARMM/ZINDO (method 1b)
Although they used a slightly different structure (PDB ID: 3ENI), 
their results for the monomer indicate similar trends as we observed here for the CHARMM/ZINDO calculations. 
In particular spectral densities have also a large overall magnitude
and the highest energy peaks are roughly at the same position where we find them for method 1b. 

\section{Summary and conclusions}
\label{conclusions}

We have modelled the thermal motion of the FMO complex for {\it C. tepidum} using two different MD forcefields. From the resulting atomic trajectories, we have then extracted 
BChl gap-energies using again two different methods, classical CDC, \rref{AdMueMa08_197_}, and quantum-chemical ZINDO/S-CIS within the ORCA package \cite{ORCA}. We can use the direct comparison of the four different sets of result to elucidate the origin of the rather large spread of literature results  \cite{OlJaLi11_8609_,OlStSc11_1771_,RiMoMa13_5510_,ShReVa12_649_,VaEiAs12_224103_} regarding the mean values of these BChl gap-energies, 
their relative ordering and the associated spectral densities. 

Regarding the distribution of energies, we find that CDC gives very narrow distributions (with width of the order of a few 10th of wavenumbers) compared to ZINDO which gives quite wide distributions (width of the order of several hundred wavenumbers). This can be understood, since a change in the bond-length within a BChl leads typically to much larger shifts of transition energies than changes in the external protein positions. In the quantum chemical method ZINDO, the former effect is fully accounted for, but not in CDC.

We confirm earlier results of \rref{OlStSc11_758_} reporting very weak inter-site energy correlations when using ZINDO, but find that correlations are much larger when using CDC. This is again consistent with a picture that energy fluctuations in CDC are dominantly determined by motion of the protein external to the BChls. These may influence several BChl due to the long range nature of interactions (and the neglect of screening). In the quantum chemical ZINDO calculations these correlations should still be present, but are there overwhelmed by the larger fluctuations due to seemingly uncorrelated internal BChl vibrations.

Also the spectral densities of CDC and ZINDO differ substantially, those from CDC having a much smaller amplitude and missing high frequency peaks,
consistent with our interpretation that these peaks come from internal vibrations. Only the ZINDO spectral densities have amplitudes of the same order as obtained from experimental FLN results \cite{RaeFr07_251_}.

We also found pronounced differences between the two forcefields.
Notably, we have obtained quite different site energies, and for the ZINDO calculations the main peak of the SD was at significantly different frequencies.
The latter finding indicates that the forcefields that we have used for the BChls result in quite different high energy vibrations.

When we averaged the SDs for a given force-field over the various temperatures and initial conditions of our ZINDO calculations, we found good agreement with experiments \cite{RaeFr07_251_} and theory \cite{VaEiAs12_224103_}, indicating rough consistency of the entire calculation scheme.  However it is the large spread in basic outcomes of the calculation without this averaging, exemplified in \fref{fig:SD_Reorgs_all}, that we consider one central result of the present work. It indicates that great care has to be taken in all details of implementing the calculation scheme.
Our interpretation of these variations, is that multiple local energy minima play a role when the BChl relax during MD structure equilibration.  In particular at low temperatures the system may stay for long times in a given minimum, but move between different configurations on even longer time scales. Jumps between certain energy values have been observed
in spectral diffusion experiments on single light harvesting complexes of various species \cite{Hofman:specdynLH2,Hofmann23122003}.

It seems that these different configurations have little influence on the SDs in the CDC approach but massively affect those in ZINDO.
An additional problem may be that some BChl configurations that are relevant for the MD are sufficiently far from the isolated equilibrium structure to cause problems for ZINDO.
 All these observations necessitate a very careful MD equilibration and choice of quantum chemical method, and point to the need for further studies for a more complete picture of the interplay of these methods.

\acknowledgments
We thank Dugan Hayes, Ulrich Kleinekath{\"o}fer,  Steffanie Valleau and Alan Aspuru-Guzik  for helpful discussions.


\begin{thebibliography}{10}
\bibitem{footnote:smoothing}
The trajectory is convolved with a Hann window function, i.e. one full period
  of a shifted cosine, with a smoothing length of $5\ \rm{ps}$.
\end{thebibliography}
\end{document}